\documentclass[preprint,times]{aastex}

\begin{document}

\title{Low Frequency VLA Observations of Abell 754: Evidence for a \\
Cluster Radio Halo and Possible Radio Relics}

\author{N. E. Kassim \altaffilmark{1}, 
T. E. Clarke \altaffilmark{2,3}, 
T. A. En{\ss}lin \altaffilmark{4}, 
A. S. Cohen \altaffilmark{1,5}
and D. M. Neumann \altaffilmark{6}}

\altaffiltext{1}{Naval Research Laboratory, Code 7213, Washington, DC, 20375 USA, kassim@rsd.nrl.navy.mil, cohen@rsd.nrl.navy.mil} \altaffiltext{2}{National
Radio Astronomy Observatory, 1003 Lopezville Dr., Socorro, NM, 87801, USA, tclarke@nrao.edu} \altaffiltext{3}{Jansky Postdoctoral Fellow} \altaffiltext{4}{Max-Planck-Institut
f{\"u}r Astrophysik, Karl-Schwarzschildstr. 1, D-85741 Garching, Germany, ensslin@MPA-Garching.MPG.DE} \altaffiltext{5}{National Research Council Postdoctoral Fellow}
\altaffiltext{6}{CEA/Saclay, Service d'Astrophysique, Orme des Merisiers, B\^at. 709, 91191 Gif-sur-Yvette Cedex, France, ddon@cea.fr}

\begin{abstract}
We present 74 MHz and 330 MHz VLA observations of Abell~754. Diffuse,
halo-like emission is detected from the center of the cluster at both
frequencies.  At 330 MHz the resolution of 90$''$ distinguishes
this extended emission from previously known point sources.  In
addition to the halo and at a much lower level, outlying steep-spectrum
emission regions straddle the cluster center and are seen only at 74
MHz. The location, morphology and spectrum of this emission are all
highly suggestive of at least one, and possibly two cluster radio
relics. Easily obtained higher resolution, higher sensitivity VLA
observations at both frequencies are required to confirm the extended
nature of the halo-like emission and the 74 MHz relic detections.
However, since there is prior evidence that this cluster is or has
recently been in the process of a major merger event, the possible
discovery of relics in this system is of great interest in light of
recent observational and theoretical evidence in favor of a
merger-relic connection. We discuss the possible role the merger shock
waves, which are seen in the X-ray emission, may have played in the
formation of the halo and radio relics in A754.
  
\vspace{0.5in} \end{abstract}


\section{INTRODUCTION}

Clusters of galaxies are the most massive gravitationally bound
objects in the universe which are in a state of quasi-equilibrium.
Our current understanding of structure formation suggests that
clusters form via merging of smaller entities, such as galaxy groups
and small clusters. During a merging process clusters are able to
release considerable energy into particle acceleration via shock waves
and turbulence \citep{deyoung92}. The accelerated relativistic
particles have a short radiative lifetime but can be reaccelerated by
recurrent merger events. For several galaxy clusters, radio
observations of the relativistic plasma reveal the presence of
extended ($>$ 500 kpc) diffuse emission which has a steep spectral
index and no observable optical counterpart \citep[and references
therein]{feretti96}.  The synchrotron nature of the emission allows us
to trace large regions of relativistic charged electrons and magnetic
fields within the X-ray emitting intracluster medium (ICM).  In
general, this diffuse emission appears to fall within two categories:
{\em halos} which are centrally located, regular in shape and
unpolarized, and {\em relics} which are located in the peripheral
regions of the cluster, irregular in shape and generally highly
polarized \citep{feretti96}.

The origin of these regions of diffuse emission is still a matter of
great debate.  The large physical scales involved combined with the
relatively short radiative lifetimes suggest that the electrons must be
re-accelerated within the ICM \citep{jaffe77}.  Theoretical models
propose that this particle acceleration may occur in the turbulence and
shocks associated with major cluster merger events \citep{deyoung92,
tribble93, feretti99, brunetti2001}, although alternative models
include secondary particle production from proton-proton interactions
\citep{dennison80}, and particle diffusion from nearby head-tail
galaxies \citep{giovannini93}. Recent work suggests that there is a
positive correlation between the presence of radio halos and relics,
and current (or recent) merger activity associated with clusters
\citep{feretti96, feretti99}.

Clusters of galaxies contain a considerable fraction of hot plasma,
the ICM, which is heated up to several $10^7-10^8$~K due to
virialization. This hot thermal plasma traces the gravitational
potential of clusters and is thus an ideal tool to determine the
dynamical state of a cluster. Because of its high temperature, which
corresponds to several keV in energy, the ICM plasma is only directly
observable with X-ray telescopes. The X-ray emission allows us not
only to trace the potential but also to trace the presence of
non-uniform heating, which is caused by adiabatic compression or shock
waves from merger events. This kind of heating, which was difficult to
detect with past X-ray telescopes is now observable in detail with the
latest X-ray satellites, such as Chandra \citep{vmm1, vmm2} or
XMM-Newton \citep{arnaud2000}.

In this paper we present new radio observations of the galaxy cluster
Abell~754 \citep{ACO}, which was observed at 74 and 330 MHz with the
NRAO VLA observatory\footnote{The National Radio Astronomy Observatory
is a facility of the National Science Foundation operated under
cooperative agreement by Associated Universities, Inc.}. X-ray
observations of this cluster \citep{1995ApJ...443L...9H, hm96},
which trace the thermal ICM, indicate that it is undergoing a
violent merger event. Recent 3D MHD/N-body simulations of
\citet{1999ApJ...518..594R, 1999ApJ...518..603R} indicate that shocks
and turbulence associated with the cluster merger could provide the
magnetic field amplification and particle re-acceleration necessary to
generate diffuse radio relic emission from regions of the
ICM. Further, the energy input from the merger may accelerate
relativistic particles in the cluster center to produce a radio
halo. Therefore A754 seems to be the ideal host for diffuse radio
emission. However, up to now radio observations have not been able to
provide an unambiguous detection of such emission. Its detection is
very important since it strongly supports the hypothesis of a
connection between radio halos/relics and merger activity.

\subsection{Past Observations of A754}

At a redshift of z = 0.054 \citep{1994AJ....107.1637B}, A754 is a
rich galaxy cluster, $\sigma_v$= 900 km ${\rm s^{-1}}$ \citep{zz95},
which has been extensively studied in the optical, radio and X-ray
bands. Observations in the 0.5-10 keV ASCA band \citep{hm96} find an
X-ray luminosity of $6.4\times 10^{44}\ h_{75}^{-2}\ {\rm ergs\
s^{-1}}$ and an emission-weighted temperature of 8.5$\pm$0.5 keV for
this cluster. Maps of the X-ray surface brightness and X-ray
temperature from ROSAT \citep{1995ApJ...443L...9H}, as well as X-ray
temperature maps from ASCA \citep{hm96} show temperature variations
that indicate that this cluster is going through or has recently gone
through a violent merger event. 

A754 has been the object of low frequency radio observations for over
20 years.  In no reported observation has a relic been detected and
evidence for a radio halo has been inconclusive.  Extended emission was
reported at 2.7 GHz \citep{1977Natur.266..239W} and at 408 MHz
\citep{1978MNRAS.185P..51M}.  In both cases the resolution of 3$'$ to
4$'$ made it difficult to distinguish apparent extended flux from the
blended images of point sources associated with galaxies in or behind
the cluster.  A higher resolution observation at 610 MHz
\citep{1980A&A....90..283H} showed no extended emission, yet the
authors conclude that a halo likely exists because, with just the point
sources they detected, they were unable to account for all emission
previously reported at lower frequencies.  This issue was later
re-investigated with the VLA in a B and C array observation (40 minutes
combined) at 330 MHz \citep{1999NewA....4..141G,2000NewA....5..335G}.
This observation detected no extended flux in the central regions of
this cluster. We note however that the short integration times obtained
at 330 MHz would provide very poor $uv$-coverage which could easily
lead to missing large scale structure in the maps.

\section{Results}

\subsection{The Data and Reduction Methods}

VLA C-array observations of A754 were conducted on March 21, 2000
simultaneously at 330 MHz and 74 MHz.  The observational parameters are
summarized in Table 1.  At both frequencies, observations were
conducted in spectral line mode to reduce the effects of bandwidth
smearing and allow for more accurate and less costly radio frequency
interference (RFI) excision.  Ultimately 8\% and 23\% of the data were
flagged at 330 MHz and 74 MHz, respectively.  Cygnus A was used as a
flux density, bandpass, and initial phase calibrator, and we estimate
flux densities we quote as accurate to $\sim15\%$.  Successive loops of
self-calibration (AIPS task CALIB) and non-coplanar, wide-field image
deconvolution (AIPS task IMAGR) were used to mitigate confusion and
achieve maximum sensitivity in our images \citep{perley99}.  The
realized rms sensitivities of $\sim6.5$ mJy/beam and $\sim200$ mJy/beam
at 330 and 74 MHz, respectively, are both approximately ten times
higher than the expected thermal noise levels. We attribute the
difference to a combination of confusion and poorly understood,
broad-band, mainly VLA-generated RFI which especially affects
low-frequency observations in the compact C and D configurations.

\begin{deluxetable}{lrr}
\tablewidth{4.5in} 
\tablecaption{Parameters of the VLA C-array Observations of A754.}
\startdata
\tableline
Frequency (MHz) & 330 & 74 \\
\tableline
Bandwidth (MHz) & 3 & 1.5 \\
\tableline
Data Type & 128 chan. & 64 chan. \\
\nodata & 1 IF & 1 IF \\
\nodata & RR\&LL & RR\&LL \\
\tableline
Time on Source (hours) & 3 & 3 \\
\tableline
Restoring Beam & $90''\times90''$ & $316''\times233''$ \\
\tableline
RMS Sensitivity (mJy/beam)  & 6.5 & 200 \\
\enddata
\end{deluxetable}

\subsection{The Radio Halo at 74 MHz and 330 MHz.}

Our 74 MHz and 330 MHz images are presented as contours in Figures
1 and 2, respectively, while a 1.4 GHz image from the NRAO VLA sky survey
(NVSS:\citet{condon}) is superimposed on Figure 1 as grey-scale.
At both frequencies we detect extended emission in the central
region of the cluster which is consistent with a radio halo, although the
5' resolution at 74 MHz is clearly too poor on its own to distinguish between 
genuine halo-emission and a blend of point source emission. However, analysis
of the 74 MHz image in conjunction with information from the 330 MHz and
1.4 GHz data indicates that there is clearly excess 74 MHz emission beyond
just a blending of the four main NVSS point sources in the halo region.
To estimate their 74 MHz contribution we determined their spectral indices
between 330 MHz and 1.4 GHz and extrapolated these to 74 MHz. The prediction
for the combined flux density of these four point sources at 74 MHz is 2.9~Jy,
while we detect 7~Jy of emission at the cluster center. Therefore we detect an
excess of about 4~Jy of 74 MHz flux in addition to the point sources. 

The case for extended emission is even stronger at 330 MHz (Figure 2)
where the resolution is sufficient to clearly separate the brightest
point sources from the diffuse emission in the region.  Apart from the
four main point sources, we measure a total of 750~mJy of flux density
in the region.  Of course it is possible that even this emission is
made up of a blending of even more faint point sources in the center
of the cluster which are undetected at 1.4 GHz, and close inspection
of the NVSS image does suggest more faint point sources in the region
(Figure 3).  However, there are regions of the halo which appear
strongly at 330 MHz but not at all at 1.4 GHz.  If we examine region A
which we define in Figure 2, the fact that it has a peak flux density
of 49 mJy/beam yet contains no observable flux at 1.4 GHz requires a
spectral index of at least 1.5, ${\rm S_{\nu}\propto \nu^{-\alpha}}$.
Taking the halo as a whole, our estimates of 4~Jy at 74 MHz and
750~mJy at 330 MHz indicate a spectral index of $\sim1.1$.  This is
unusual for point sources associated with radio galaxies.  The four
brighter point sources all have spectral indices ranging from 0.5 to
1.0. We therefore assert that while some of this emission could be due
to point sources, a significant fraction corresponds to a diffuse
cluster radio halo with 3$\sigma$ major and minor axes of
$\sim450''\times400''$, respectively. At the redshift of the cluster
this corresponds to a scale of $\sim430 \times 380~h_{75}^{-1}$~kpc, a
factor of 2-3 smaller than the largest known giant Mpc halos but still
not the smallest halo known \citep{feretti99}. From our 330 MHz image
we estimate the halo center at $\sim09^h08^m43^s$,$-09^\circ36'57''$,
indicating it is shifted westward of the X-ray maximum and also of the
cluster center, which we place at
$\sim09^h09^m03^s$,$-09^\circ39'46''$ with reference to an elliptical
beta-model fit.  We note that the $\alpha$=1.1 spectrum of the diffuse emission
we infer is consistent with previous estimates
\citep{1980A&A....90..283H} though the total flux levels are still
somewhat ($\sim40\%$) lower than predicted.

Initially puzzling, 330 MHz observations by \citet{2000NewA....5..335G}
at comparable rms noise failed to detect the halo emission. However, a 
plausible explanation emerges in light of the fundamentally different 
spatial characteristics of the two data sets. Our observations were 3 
hours in the VLA C array, while theirs was a combination of B and C array 
snapshots (40 min total) with consequently less thorough $uv$ coverage. 
A possible explanation is therefore that their superior resolution 
provided by B-array baselines sets a lower basic confusion level and 
thus provided a nominal sensitivity comparable to our longer, lower 
resolution integration.  However, their snapshot C array $uv$ coverage 
would at the same time have been much less sensitive to the large-scale 
extended emission we detect.
Indeed imaging of a snapshot sub-section of our own C array data failed
to detect the halo.

\subsection{Discovery of Radio Relics at 74 MHz.}

Our 74 MHz image (Figure 1) reveals two, newly detected objects without
counterparts at either 330 MHz or 1.4 GHz which we label as the ``East
Relic'' and ``West Relic'' (see Table 2). While the $6\sigma$
(integrated flux $\sim1.45$ Jy) and $4\sigma$ ($\sim850$ mJy,
unresolved) detections are weak, it is nevertheless highly unusual to
detect sources at 74 MHz which are not also seen by the VLA at either
330 MHz or 1.4 GHz. In fact {\em all} other objects in a 4$^\circ
\times {\rm 4^\circ}$ 74 MHz image that appear at the brightness of the
fainter West Relic or higher have clear counterparts in the 1.4 GHz NVSS images, and 330 MHz counterparts within the $\sim2^\circ$ radius region in which the 74 and 330 MHz fields of view overlap. This suggests an unusually high spectral
index for these objects.  In order to explain the lack of a detection
at above three times the rms sensitivity in the 330 MHz or 1.4 GHz
images, the ``East Relic'' would need a spectral index of at least 1.8
or 1.4 respectively.  Such high spectral indices are typical of cluster
radio relics.

\begin{deluxetable}{llll}
\tablewidth{4.5in} 
\tablecaption{New Objects in A754 at 74 MHz}
\startdata

\tableline
Object & Peak & \multicolumn{2}{l}{Position (J2000)} \\
ID & Jy/beam & RA & DEC \\
\tableline
East Relic & 1.25 & $09^h09^m36^s$ & $-09^\circ43'22''$ \\
\tableline
West Relic & 0.85 & $09^h07^m25^s$ & $-09^\circ36'52''$ \\

\enddata
\end{deluxetable}

There is further evidence that these objects detected at 74 MHz are
indeed radio relics.  First the morphology of the ``East Relic'', with
an elongation perpendicular to its separation from the cluster X-ray
maximum is highly suggestive of a relic. Its major and minor axes
are $\sim600''\times230''$, respectively, which corresponds to scale
sizes of $\sim580 \times 220~h_{75}^{-1}$~kpc at the redshift of the
cluster. The reality of the mainly unresolved ``West Relic'' is clearly
much more speculative, though it is noteworthy that its location with
respect to the cluster X-ray maximum is opposite to that of the
``East Relic''. An overlay of the positions of these relics with
ROSAT hard X-ray data (Figure 4) reveals that the East relic falls
exactly at the location of flattened, compressed X-ray contours. Such a
rapid gradient in X-ray surface brightness is expected at the location
of merger shock waves.  Further, \citet{1998ApJ...493...62R} have
predicted locations of shock waves based on their model of the cluster
merger in A754, and it is especially noteworthy that the location of
the East relic coincides with a jump in the X-ray temperature
predicted for the shock waves in this cluster. We note however that
the signal-to-noise ratio on both of these new objects is still very
low and thus the objects (the West relic in particular) should be
regarded with caution. We plan to undertake deep follow-up observations
of these sources with the VLA at higher resolution to confirm
these detections and provide tighter constraints on their spectra.
Additional follow-up observations will be undertaken to identify the
compact X-ray source located at the position of the East relic and
determine if the two objects are connected.

\section{Discussion}
\label{disc}

In the numerical model of \citet{1998ApJ...493...62R} a large
subcluster (cluster to subcluster mass ratio 2.5:1) has passed from
East to West slightly off-axis from the core of the A754 system. This
merger event results in the formation of two shock waves which are
visible as temperature jumps in Fig. 2a of their work. The Eastern
shock heats the infalling, stripped subcluster gas in a relatively
small area. The location of this shock at the ram pressure flattened
edge of the X-ray peak is coincident with the observed location of the
East relic source.  The western shock heats the outer regions of the
main cluster ICM and is much more extended. In principle, this shock
could have caused the West relic, but we note that the West relic
is further away from A754's X-ray peak ($\approx 1.8\,h_{75}^{-1}\,
{\rm Mpc}$) than the shock wave in the numerical model ($\approx 1.2\,
h_{75}^{-1}\, {\rm Mpc}$).   The measured \citep{1995ApJ...443L...9H,
hm96} and simulated \citep{1998ApJ...493...62R} temperature maps show
that mainly the outer western region is heated by this shock wave, but
not the innermost cluster core region.

The temperature variation suggests that the impact of the infalling
gas combined with the change in the gravitational potential of the
primary cluster due to the core passage of the dark matter subcluster
was not violent enough to produce a shock wave in the dense main
cluster core. The core was only compressed adiabatically, and relaxed
nearly to its original temperature in a later merger stage. When the
impact moved into outer, less dense regions, a shock wave was created
and heated the gas non-adiabatically. Such emergence of outgoing shock
waves is observed in numerical simulations of cluster merger events
(Schindler, private communication). 

A fraction of the energy dissipated in shock waves is often transfered
into relativistic particle populations. Accelerated relativistic
electrons have short radiative lifetimes ($\sim 0.1$ Gyr) and
should therefore produce the radio relic emission close to the current
location of the shock waves.  The electron acceleration required to
produce the relic emission could result from Fermi-I diffusive shock
acceleration of ICM electrons \citep{1998A&A...332..395E,
1999ApJ...518..603R}, or adiabatic energization of relativistic
electrons confined in fossil radio plasma, released by a former active
radio galaxy \citep{2001A&A...366...26E}.

Observationally, the radio halo of A754 appears to be displaced from
the main cluster core in the direction of the West relic. This
suggests that the halo emission results only from regions which were
heated by the re-emerging western shock wave after core passage.
Since the radiative cooling time of the radio halo electrons
observed at 330 MHz is smaller than 0.3 Gyr, but the shock passage was
more than 0.3 Gyrs ago (at least in the numerical model), the radio
electrons must have been re-populated after shock passage. In situ
Fermi II re-acceleration (powered by some residual plasma turbulence
after the shock passage) of shock accelerated electrons could be able
to re-accelerate the electrons \citep{brunetti2001}, if the turbulence
is able to remain for a few 100 Myr. Alternatively, relativistic
protons might be efficiently accelerated at shock waves and would have
lifetimes comparable to the cluster age. Since these protons can seed
relativistic electrons and positrons into the ICM by hadronic
collisions with the ICM ions, they might be responsible for the radio
halo emission which glows for a long time after the shock passage
\citep{dennison80, 1982AJ.....87.1266V, 1999APh....12..169B,
2000A&A...362..151D}. This process would only be efficient in the
dense cluster core regions and thus is likely not important to the
formation of peripheral radio relics.

The above scenarios are speculative but make clear predictions which
further observations can confirm. First, further observations can
reveal that either or both of the relic structures are real. Second, we
expect a high degree of source-intrinsic radio polarization in the
East relic, due to the nearly perpendicular line of sight to the
shock compression direction. The compressed magnetic fields should be
roughly aligned with the shock plane and relic major axis, thus the
synchrotron emission should be highly polarized. For the West relic,
the geometry is less constrained, but polarization is also very
likely.

Further, the steep spectral index of the East relic can be
understood, but would likely have different spectral curvature, in the
two relic formation models. In the Fermi I shock acceleration model a
low Mach number shock would produce a steep power-law radio spectrum
(Mach number $<3.5$ for a spectral index steeper than 1.8). In the
revived fossil radio plasma model the steep spectrum would be a
synchrotron/inverse Compton cooling cut off, which was shifted by the
shock compression close to the observed frequency.
Additional information on the spectral curvature of the
relics might therefore help to discriminate these two scenarios.

Finally, if radio halos occur only in the shocked ICM -- as A754
seems to indicate --, then one would expect a spatial correlation
of shock-heated gas and diffuse synchrotron emission in other clusters
which are in the early stages of major merger events. Such a
correlation could be observable with high-resolution observations from
X-ray satellites such as Chandra. At later merger stages, the
turbulent gas motion, which is stirred by the {\it violently relaxing}
dark matter cores, should have erased many of the temperature and
radio halo substructures.

The hypothesis that radio halos occur only in shocked ICM
could explain why the post merger cluster A3667 has no radio halo,
despite the fact that two giant cluster radio relics
\citep{1997MNRAS.290..577R} indicate the presence of peripheral
merger shock waves \citep{1998A&A...332..395E}. In this case the main
cluster core seems to be unshocked \citep[see temperature map in
their Fig. 3]{1999ApJ...518..603R} because the merger was likely less
violent. There only a mass ratio of the merging clusters of 5:1 is
needed in order to reproduce its X-ray morphology, compared to 2.5:1
for A754 \citep{1998ApJ...493...62R}. In the re-acceleration scenario
\citep{brunetti2001}, the shock waves would have either injected
energetic electrons only in peripheral, low magnetic field regions, so
that no radio halo occurs, or the necessary re-acceleration turbulence
has decayed in the late merger state of A3667 (1 Gyr after core
passage). The re-acceleration model can be tested due to the high
energy cutoff expected in the radio electron spectrum at energies
where the cooling time-scale is shorter than the acceleration
time-scale. A stronger halo suppression can be expected in the
scenario of radio halo electrons being produced by shock accelerated
protons. In this scenario the shock waves in A3367 would have emerged at radii
which were too peripheral to host a sufficiently dense thermal proton
population required as targets for the hadronic secondary electron
production to operate efficiently -- in addition to the weaker radio
emissivities there due to weaker magnetic fields .  The hadronic halo
scenario will be tested with the next generation Gamma ray telescopes
for the unavoidable Gamma radiation from neutral pion decays following
hadronic proton proton collisions \citep{1982AJ.....87.1266V,
ensslin97, 1998APh.....9..227C, 1999ApJ...525..603B,
2000A&A...362..151D}.

\section{Conclusions}

A754 has been successfully imaged at both 330 MHz and 74 MHz.  We
conclude these images strongly suggest the existence of both a radio
halo and at least one and possibly two radio relics in this galaxy
cluster.  This finding is especially relevant in light of recent
observational and theoretical evidence in favor of a merger-relic
connection.  Deeper 74 and 330 MHz imaging at higher resolution are
clearly warranted to confirm the extended nature of the radio halo and
the relic detections. These observations suggest that further study of
this system and similar ones at low frequencies will open a new window
of understanding on cluster dynamics and cluster formation.

\section{Acknowledgments}

The authors would like to thank Kurt Roettiger for extensive help with the
observations and at early stages of the project. We also thank Rick
Perley and Frazer Owen for useful discussions, as well as our anonymous
referee whose comments and suggestions improved our paper. NEK also
thanks D. Harris and L.  Feretti for the Ringberg Castle invitation
which inspired the project.  We have made use of the ROSAT Data Archive
of the Max-Planck-Institut f\"ur extraterrestrische Physik (MPE) at
Garching, Germany.  Basic research in radio astronomy at the Naval
Research Laboratory is funded by the Office of Naval Research.

\newpage

\begin{figure}
\vspace{6in}
\includegraphics{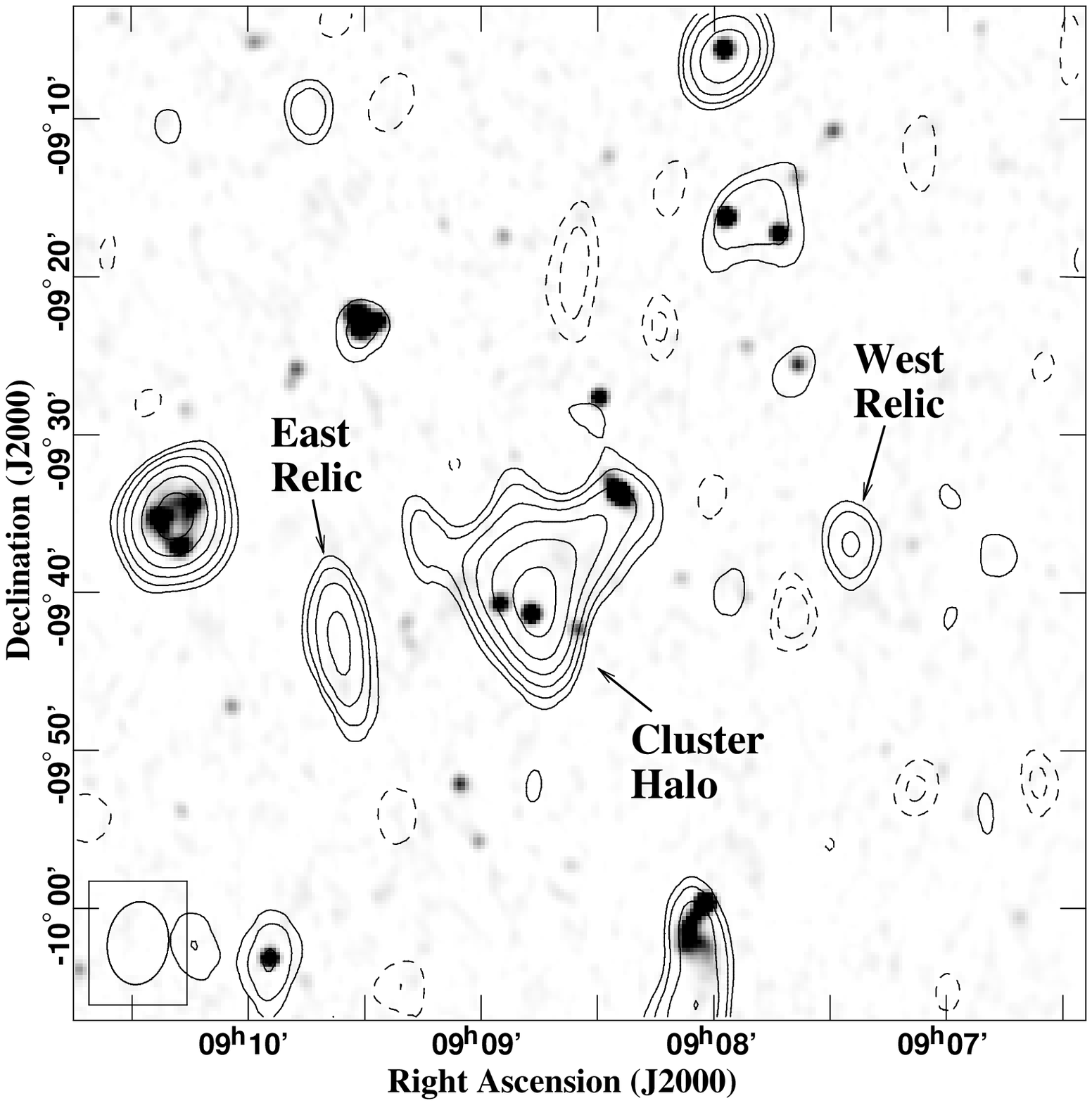}
\caption{The contours show a 74 MHz VLA C-array image of A754.  We
identify the cluster radio halo and two possible radio relics. For
comparison, we show in grey-scale the NVSS 1.4 GHz image as well.  For
the 74 MHz image, the peak flux density is 4.2 Jy/beam and the beam
size is $316'' \times 233''$ at position angle of
$-7^{\circ}$.  Contours levels are 0.4 Jy/beam $\times (-1.4, -1, 1,
1.4, 2, 2.8, 4, 5.7, 8, 11.3)$.  The off-source RMS noise is 0.2
Jy/beam.  }
\end{figure}

\newpage

\begin{figure}
\vspace{6in}
\includegraphics{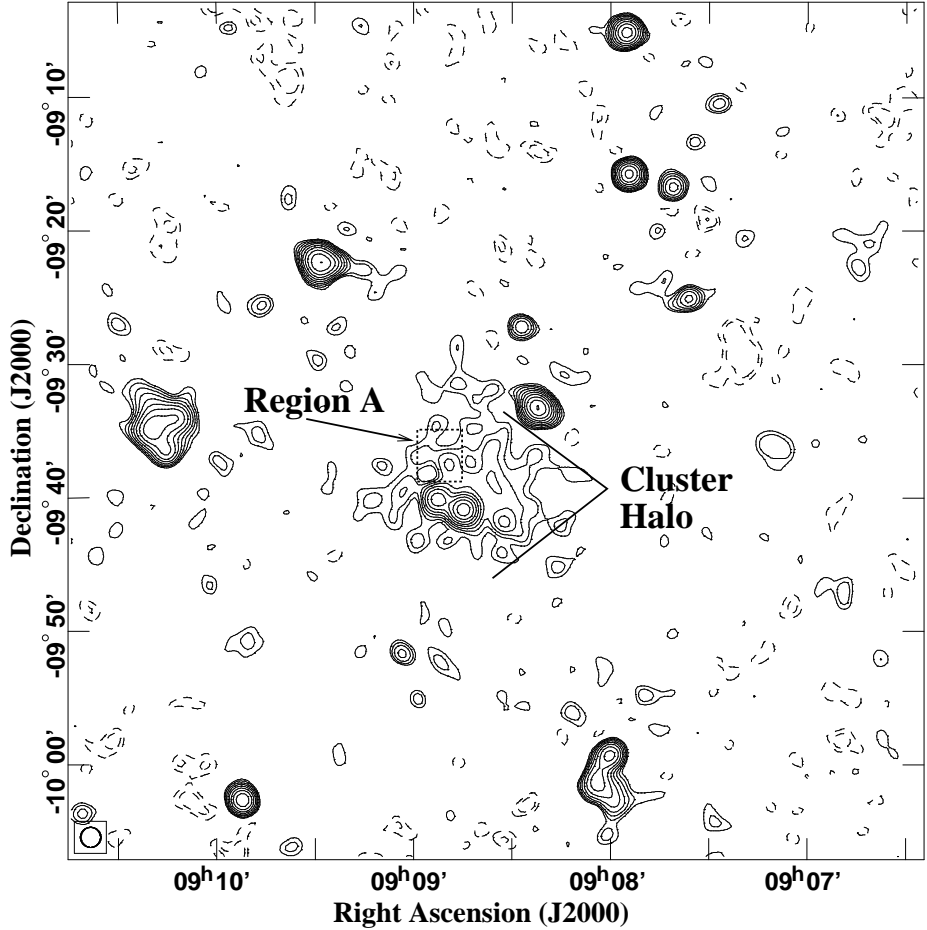}
\caption{A754 at 330 MHz.  This image was taken with the VLA in
C-array and clearly shows what appears to be a radio halo.  Also
visible are four bright points sources in the halo region which also
appear at 1.4 GHz.  The spatial scale is identical to Figure 1.  The
peak flux density is 567 mJy/beam and the restoring beam size is
$90'' \times 90''$.  Contours levels are 16
mJy/beam $\times (-1.4, -1, 1, 1.4, 2, 2.8, 4, 5.7, 8, 11.3, 16, 22.6,
32)$.  The off-source RMS noise is 6.5 mJy/beam.  }
\end{figure}

\newpage

\begin{figure}
\vspace{6in}
\includegraphics{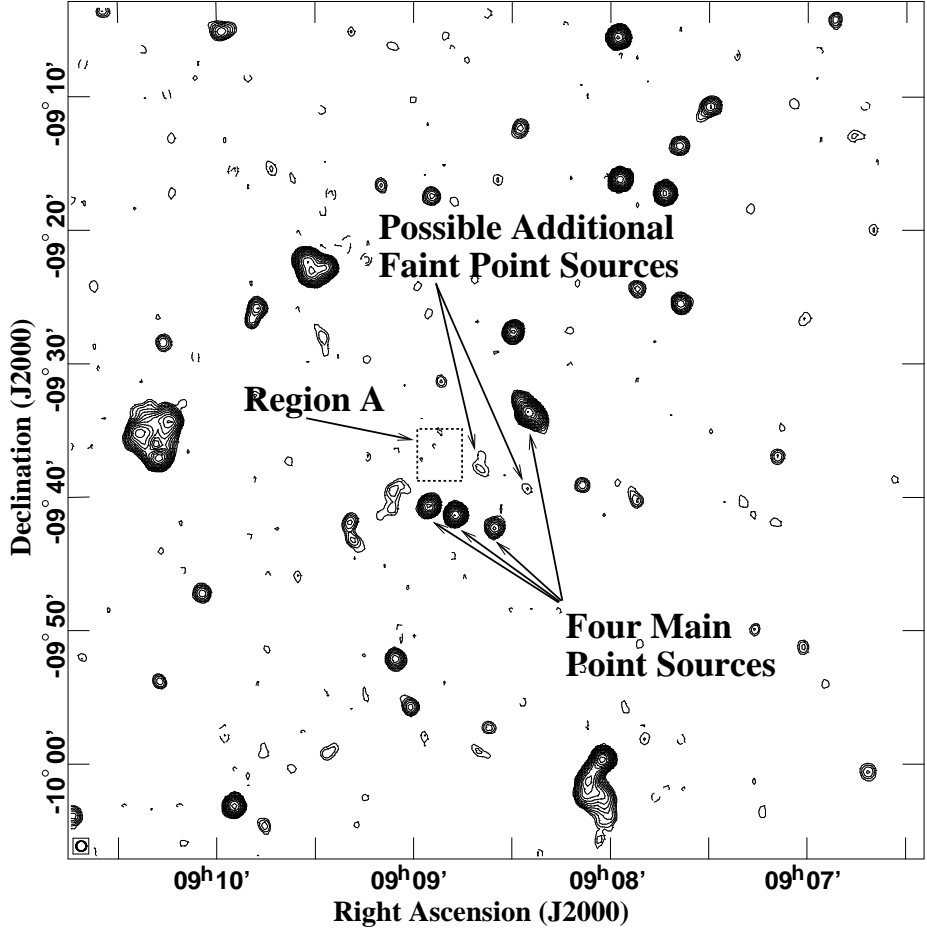}
\caption{A754 at 1.4 GHz.  This image was created from the NVSS catalog 
and is shown at the same spatial scale as Figures 1 and 2.  The 
point sources here provide a reference for comparison with the 330 MHz 
image.  The peak flux density is 181 mJy/beam and the restoring beam 
size is $45'' \times 45''$.  Contours levels are 1.4 
mJy/beam $\times (-1.4, -1, 1, 1.4, 2, 2.8, 4, 5.7, 8, 11.3, 16, 22.6, 
32, 45.3, 64, 90.5, 128)$.  The off-source RMS noise 
is 0.45 mJy/beam.}
\end{figure}

\newpage

\begin{figure}
\vspace{7in}
\hspace{-0.4in}
\includegraphics{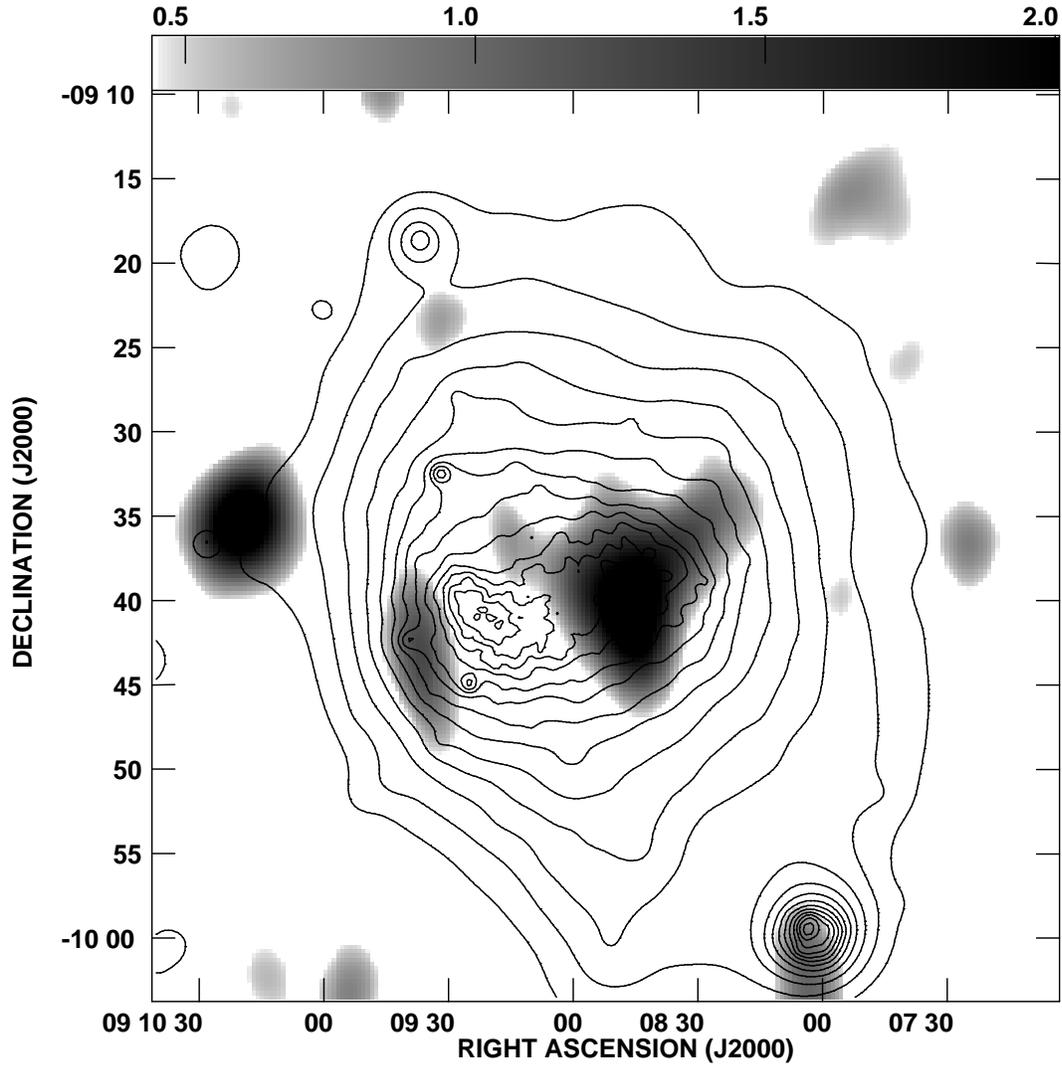}
\caption{Overlay of 74 MHz VLA data in greyscale with the ROSAT
0.1-2.4 keV X-ray contours. The location of the East relic is
co-incident with the ram pressure compressed X-ray contours. Further,
the radio halo and relic emission appear to avoid the dense X-ray
bright region of the cluster.  We note that there
appears to be a compact X-ray source at the location of the East relic
but as yet there is no optical identification for this object.}
\end{figure}

\end{document}